# Recent Breakthrough in AI-Driven Materials Science: Tech Giants Introduce Groundbreaking Models

Miao Liu, Sheng Meng

Institute of Physics, Chinese Academy of Sciences, Beijing 100190, China

Songshan Lake Materials Laboratory, Dongguan, Guangdong 523808, China

Email: mliu@iphy.ac.cn; smeng@iphy.ac.cn

In late November 2023, DeepMind, a subsidiary of Google, made a groundbreaking announcement in the prestigious journal Nature[1]. They unveiled a powerful artificial intelligence (AI) model designed for materials science, named Graph Networks for Materials Exploration (GNoME). Through the application of GNoME and high-throughput quantum-mechanics-based first-principles calculations, researchers at DeepMind discovered over 384,781 thermodynamically stable crystalline materials, which is equivalent to "nearly 800 years' worth of knowledge" as quoted from their website, drastically accelerating the pace of new materials discovery.

Mere days after Google announced the GNoME model, in December 2023, Microsoft released its own AI endeavor in materials science - the generative model MatterGen[2], which can predict new material structures based on targeted properties. Taking to social media, Microsoft's President lauded the company's AI model, commenting, "With our new MatterGen model, we're applying the next generation of AI to one of the biggest challenges in materials science: increasing the rate at which we design materials with desired properties."

In a significant partnership move in January 2024, Microsoft collaborated with the Pacific Northwest National Laboratory (PNNL), under the U.S. Department of Energy, to further advance the field. Utilizing a combination of AI and high-performance computing, they filtered through 32 million inorganic materials, which were created by a team at the University of California, to identify novel solid-state electrolytes.[3] They indeed found one and closed the loop from theoretical prediction to experimental validation.

Materials science unequivocally stands as a cornerstone of modern industry. The early stages of human civilization—Stone, Bronze, and Iron Ages—are intimately intertwined with material progression. The invention of ceramics laid the foundation for the prosperity of ancient China, while glassmaking catalyzed the development of optical devices, setting the stage for breakthroughs in cellular biology and astronomy. One could argue that the history of human civilization is intrinsically a history of the evolution of materials science.

The rapid developments from both Google's and Microsoft's temptations signal that we are in a new transformative stage in materials science. These AI-driven approaches are expediting the discovery of new materials to meet industrial needs. Aside from Google and Microsoft, Meta and ByteDance have also been laying out similar research and development directions. For a time, tech giants have created a surge of activity in a new racetrack, using all the technologies that have disrupted other business sectors and started to disrupt materials science.

How is AI revolutionizing materials R&D? The industry giants have identified a common technological trajectory: (1) acquiring materials science data through theoretical calculations; (2) producing massive amounts of such data via high-throughput computing; (3) feeding this data to AI models; (4) using the AI models to infer the properties in undiscovered materials. They all foresee that this is a fundable direction with a broad and promising outlook.

Will AI revolutionize the way materials science research is conducted in the future? The short answer is yes. The recent advances in information technology have already reshaped many behaviors of our society. Data, algorithms, and computational power become the three key factors driving this transformation.

Withstanding this frenzy, let us take a closer look at this very progress released by Google, delving into its contents to understand its significance, scope, and possible implications for the future of materials science. Here are 11 things you need to know:

1. Materials science has become the next significant area where AI is making substantial inroads. AI has been successfully applied to various businesses and applications, from analyzing medical images to autonomous driving vehicles. Data catalyzes AI's takeoff. Thanks to the development of the Materials Genome Initiative and various materials science databases, this field now possesses high-quality data resources, which are prerequisite conditions for the rise of AI.

2. Datasets sets the foundation for the development of AI models. The AI field has a high dependence on data, and the scope and quality of datasets directly dictate the capabilities of AI models. The breadth of a dataset determines the model's generalizability, while consistency and comparability determine its predictive accuracy. Among the three key elements of AI—data, algorithms, and computing power—data is the most critical and challenging barrier. Even if model source codes are open-sourced, without a good dataset, it is challenging to properly train good models.

3. Theoretical calculations have contributed significantly to the creation of materials science databases. Density functional theory (DFT), after decades of development, now has a mature reservoir of techniques that can produce highly standardized datasets in a short period. For example, the most widely used dataset in the field of materials science, Materials Project[4], is obtained through DFT calculations. According to the current state of materials science, it is less likely to achieve similar-level datasets via experimental means.

4. The content released in Google's paper includes both the GNoME model code and the dataset. The dataset boasts high coverage in chemical space with good precision. The dataset was derived from the Materials Project[4] and OQMD[5], and adopts the same computational standards and workflows, allowing for its integration with the Materials Project. Google claims that it has created a database of 2.2 million inorganic materials using high-throughput computing and DFT, continually predicting new thermodynamically stable materials through active learning and eventually identifying 384,781 stable inorganic compounds, giving a considerable boost to the field of materials science.

5. Although Google holds a large GNoME dataset encompassing 2.2 million inorganic materials, the information released with the paper includes only the structures, thermodynamic stability, and source code of the AI models. Google has not made public the model parameters; thus, third parties are unable to run the model as an off-the-shelf solution. Additionally, Google has not released sufficient data for effective model training, making Google the sole holder of the GNoME model.

6. We conducted a detailed analysis of the structural information of the 384,781 compounds. We found that the elemental combinations of 30,345 materials (for example, "Zr-Ti-Se", and "Ni-Te") could be found in the Materials Project database, making up 7.8% of the GNoME dataset (Figure 1). This means that within the chemical space that has been explored by humans, Google has identified 30,345 thermodynamically stable materials. The majority (92.2%) of the stable materials come from elemental combinations yet unexplored by humans (for instance, "Rh-Ac", and "Zn-Cs"). This suggests that many undiscovered stable compounds remain within the unknown chemical space, and the materials known to people so far might just be the tip of the iceberg. However, for the most part, these stable materials from uncharted chemical spaces contain elements in low abundance, which casts doubt on their practical application value.

7. Google's GNoME model samples from a broader chemical space, which likely gives it a unique advantage in generalization capability compared to other AI models[6,7] typically developed using data from the Materials Project alone. (Figure 1)

8. In the GNoME dataset from Google, metallic materials constitute over 60% of the entire database(Figure 1). The presence of many unknown stable structures within metallic materials is reasonable because metal elements tend to bind to form metallic bonds, thereby lowering the system's energy. However, these metal elements often form alloys with randomly distributed sites rather than the intermetallic compounds as present in the GNoME dataset, which means they are unlikely to be synthesized as predicted. Nevertheless, these data are still very much meaningful for the training of AI models.

9. The frequency of element occurrences in Google's GNoME dataset differs significantly from that in the Materials Project (Figure 2). There is small percentage of ionic compounds in the GNoME dataset as there is a higher occurrence of metal elements, especially those that are less abundant, such as Ho, Tb, Rh, and Er, whereas common elements like O, P, and S appear less frequently.

10. Google's GNoME dataset contains a large proportion of multinary metallic compounds and doped structures (Figure 3), which are typically difficult to synthesize. A statistic seen in Figure 3 shows the atomic percentage of the minority elements in the compounds relative. Many stable binary compounds are, in fact, doped phases. For example, $HBr_{35}$ is a structure with one H atom in 35 Br atoms, making it challenging to arrange H periodically and regularly within this structure to form a crystal that is consistent with the predicted stable structure. For ternary and quaternary phases, the doped structures appear less frequently.

11. The advancement of AI algorithms found in visual and language models will undoubtedly find their applications in materials science soon. Reinforcement learning, attention mechanisms, diffusion models, pre-trained models, multimodal technologies, generative algorithms, model alignment mechanisms, vector databases, and more will inevitably be continuously introduced into materials science, producing corresponding tools.

The dataset mentioned by Google is a testament to this transformative era. While specifics of the dataset are not disclosed, it can be assumed to encompass intricate details concerning the 384,781 newly identified materials. Such a dataset opens a new realm of possibilities. Researchers across the globe will have the opportunity to explore these materials further, potentially uncovering novel applications and optimizing existing ones. It is more than just a dataset; it is a roadmap to untold innovations that could reshape the world.

4. Jain, A. *et al.* Commentary: The materials project: A materials genome approach to accelerating materials innovation. *APL Materials* **1**, 1 (2013).
5. Saal, J. E., Kirklin, S., Aykol, M., Meredig, B. & Wolverton, C. Materials design and discovery with high-throughput density functional theory: The open quantum materials database (OQMD). *JOM* **65**, 1501–1509 (2013).
6. Chen, C. & Ong, S. P. A universal graph deep learning interatomic potential for the periodic table. *Nat Comput Sci* **2**, 718–728 (2022).
7. Deng, B. *et al.* CHGNet as a pretrained universal neural network potential for charge-informed atomistic modelling. *Nat Mach Intell* **5**, 1031–1041 (2023).

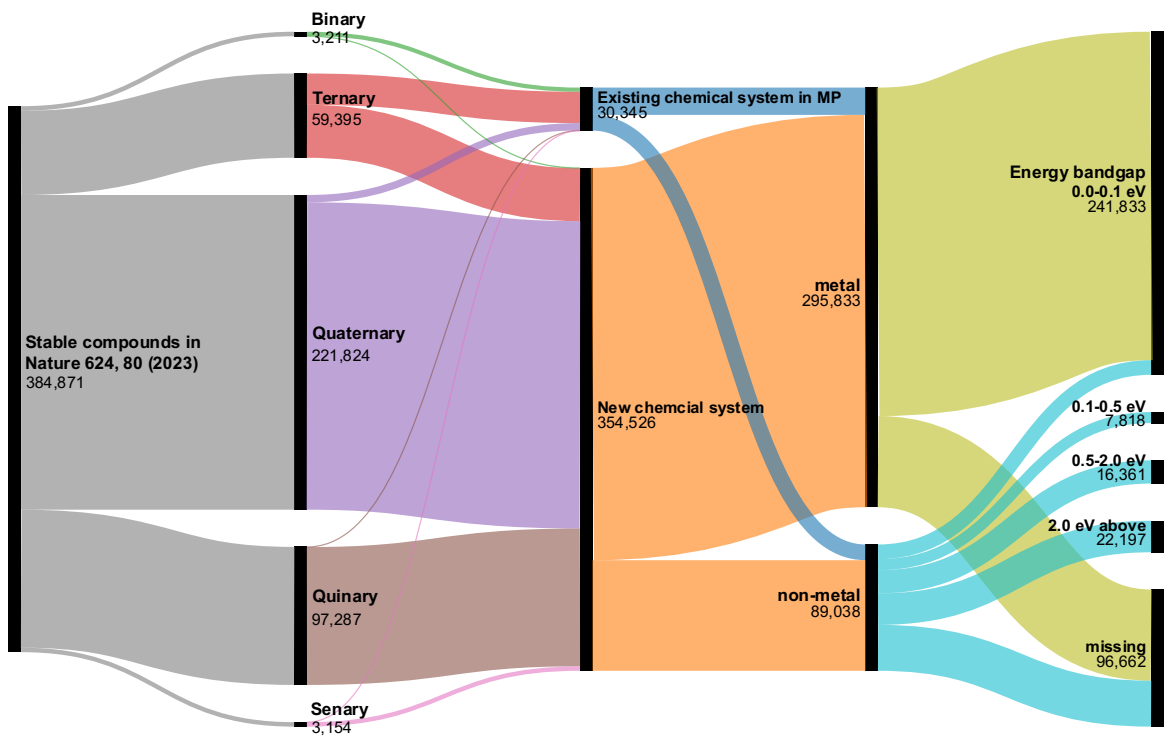

Figure 1. A close look at the GNoME dataset on its chemical system, metallicity, and energy band gap.

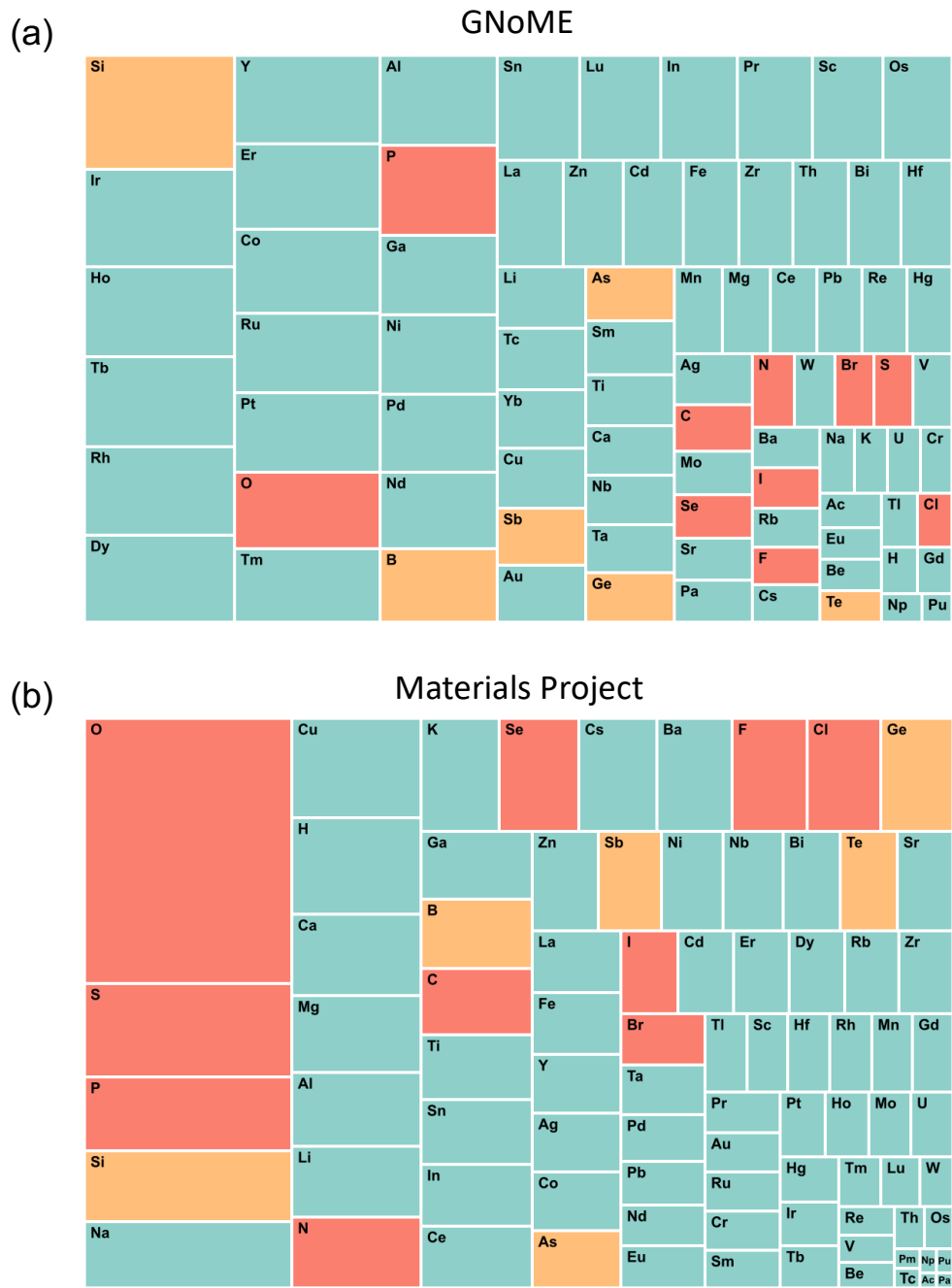

Figure 2. The likelihood of finding a certain atom species in both GNoME and Materials Project.

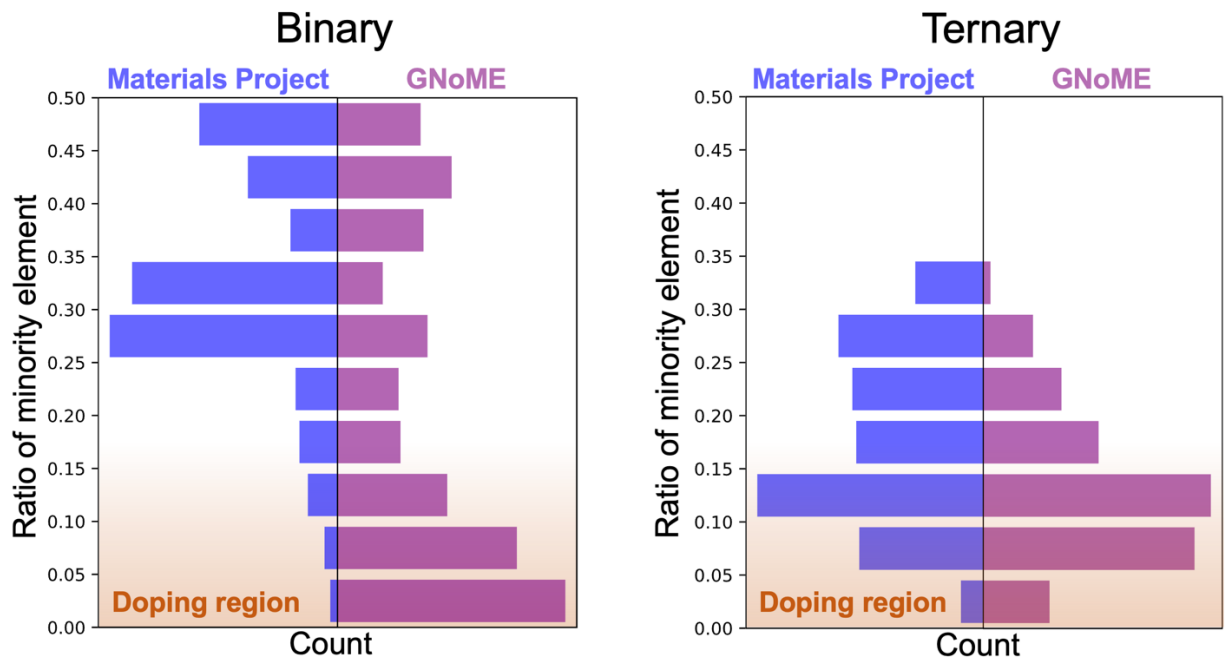

Figure 3. The statistic on the ratio of the minority element in a binary or ternary compound in the GNoME dataset.